\documentclass[
aps,
prb,
twocolumn,
superscriptaddress,
longbibliography
]{revtex4-2}

\usepackage{amsmath,amssymb,bm}
\usepackage{braket}

\usepackage{graphicx}
\usepackage{subfigure}

\usepackage[
colorlinks=true,
linkcolor=blue,
citecolor=blue,
urlcolor=blue
]{hyperref}
\begin{document}
	
	\title{Asymmetric quantum information scrambling in inhomogeneous XXZ spin chains}
	\author{Shivam~Mishra}
	\email{shivam.2023rph01@mnnit.ac.in}
	
	\author{Ravi~Prakash}
	\email{ravi.prakash@mnnit.ac.in}
	
	\affiliation{Department of Physics, Motilal Nehru National Institute of Technology Allahabad, Prayagraj--211004, India}
	
	\begin{abstract}
		Inhomogeneous systems have become increasingly relevant in experimentally engineered quantum many-body systems, where spatially varying interaction strengths can significantly influence nonequilibrium dynamics. Motivated by this, we numerically study the out-of-time-ordered correlators (OTOCs) in inhomogeneous XXZ spin chains. Using two types of spatially varying interaction profiles, namely, linear-gradient and stepwise profiles, we show that spatial inhomogeneity in the interactions can induce pronounced asymmetry in information scrambling, with OTOC operators located at different sites exhibiting distinct dynamical behavior. In particular, OTOCs associated with the strongly interacting side exhibit suppressed scrambling compared to those associated with the weakly interacting side, with the difference becoming more pronounced when the interaction strengths differ significantly. To elucidate the origin of the long-time saturation of OTOCs, we analyze the diagonal matrix elements of the OTOC observables in the energy eigenbasis and their overlap with the Hamiltonian, which explicitly incorporates the spatial interaction profile. The resulting analytical expression is consistent with the numerical results. Finally, we extend our analysis to periodically driven inhomogeneous spin chains and show that the interaction-gradient-induced asymmetry persists under periodic driving.
	\end{abstract}

	\maketitle
	
	\section{Introduction}
	Understanding the propagation of quantum information in isolated quantum many-body systems has become a central problem in nonequilibrium physics. In interacting systems, information initially encoded in local operators spreads under unitary time evolution and becomes increasingly nonlocal \cite{int_1,int_2}. A powerful probe of this phenomenon is provided by out-of-time-ordered correlators (OTOCs) \cite{int_3,int_4,int_5,int_6,int_7,int_8,int_9,int_10,int_11,int_12}, which characterize operator growth and scrambling dynamics \cite{int_1,int_8,int_13,int_14}. OTOCs have attracted much attention recently because of their promising applications and experimental accessibility for diagnosing information scrambling in the dynamics of quantum many-body systems \cite{exp_15,exp_16,exp_17,exp_18,exp_19,exp_20,exp_21,exp_22,exp_23}.
	
	For two initially separated local unitary operators $\hat W$ and $\hat V$, the OTOC is defined as the expectation value of their squared commutator, $C(t)=\frac{1}{2}\langle [W(t),V]^\dagger [ W(t), V]\rangle$, which is often rewritten in terms of the corresponding OTO correlator $\mathcal{F}(t)=\langle W^\dagger(t) V^\dagger  W(t) V\rangle$ such that $C(t)=1-\mathrm{Re}[\mathcal{F}(t)]$. Throughout this work, we consider the infinite-temperature regime, where the expectation value is evaluated with respect to the maximally mixed density matrix $\rho=\mathbb{I}/2^L$, so that $\langle \cdots \rangle=\mathrm{Tr}(\rho \cdots)$. Initially, the two local operators commute because they are spatially separated. However, under unitary time evolution, the operator $W(t)$ spreads across the system and develops overlap with $V$, causing the OTO commutator to grow. 
	In systems with local interactions, the spreading is constrained by the Lieb-Robinson bound, leading to an effective light-cone structure for the propagation of quantum information \cite{Lib_rob1,Lib_rob2}.
	
	OTOCs have been extensively studied in systems ranging from black-hole physics \cite{int_4,int_6,black_hole,black_hole1} and Sachdev--Ye--Kitaev (SYK) models \cite{syk,int_11,syk2,syk3} to quantum spin chains \cite{spin_chain,spin_chain1,spin_chain2,spin_chain3,spin_chain4,spin_chain5,spin_chain6,spin_chain7,spin_chain8,spin_chain9}, encompassing both chaotic and integrable regimes. Considerable effort has also been devoted to developing protocols for measuring OTOCs \cite{int_8,experiment1,experiment2}. Recent advances have enabled their experimental realization using nuclear spins in molecules \cite{nmr}, trapped ions \cite{trapped_ions}, and ultracold atomic gases \cite{ultra_cold}.
	The OTOC has emerged as a prominent measure of information scrambling in many-body systems. However, growing evidence suggests that the notion of chaos captured by OTOCs does not necessarily coincide with that inferred from conventional spectral diagnostics \cite{Balachandran2021,Balachandran2023,Pappalardi2018,Xu2020Chaos}. The OTOC of a generic system is expected to decay rapidly, with the decay rate carrying information about the underlying chaotic properties of the system, and to approach a long-time saturation value. A saturation value close to zero indicates nearly complete information scrambling, whereas a finite saturation value points to restricted scrambling \cite{unscrambling}. It is well known that the initial power-law growth of the OTOC is a generic short-time property of spin chains and scales as $t^{2d}$, where $d$ is the distance between the OTOC operators, and is independent of the Hamiltonian parameters~\cite{spin_chain3,early_time1,early_time2}. 
	
	The XXZ spin chain provides a natural platform for this study owing to its well-established role in investigations of quantum chaos, thermalization, transport, and nonequilibrium dynamics. Deterministic spatially varying interactions introduce a controlled, disorder-free route to integrability breaking (see, e.g., Refs.~\cite{hamiltonian,gaurav_krylov}), enabling the system to evolve from an integrable regime to quantum chaos and, at sufficiently large interaction gradients, toward a nonergodic regime. Recent advances in programmable Rydberg-atom arrays have demonstrated controlled Hamiltonian engineering through spatially tunable spin interactions \cite{exp_inhomo}, providing an experimental route to realizing deterministic interaction gradients. Such interaction gradients are expected to induce directional asymmetry in information scrambling. Recently, it has been numerically shown that increasing spatial inhomogeneity can suppress information scrambling and ergodic dynamics in finite-sized quantum spin systems~\cite{hamiltonian,gaurav_krylov}. 
	
	\begin{figure}[t]
		\centering
		\includegraphics[width=0.48\textwidth]{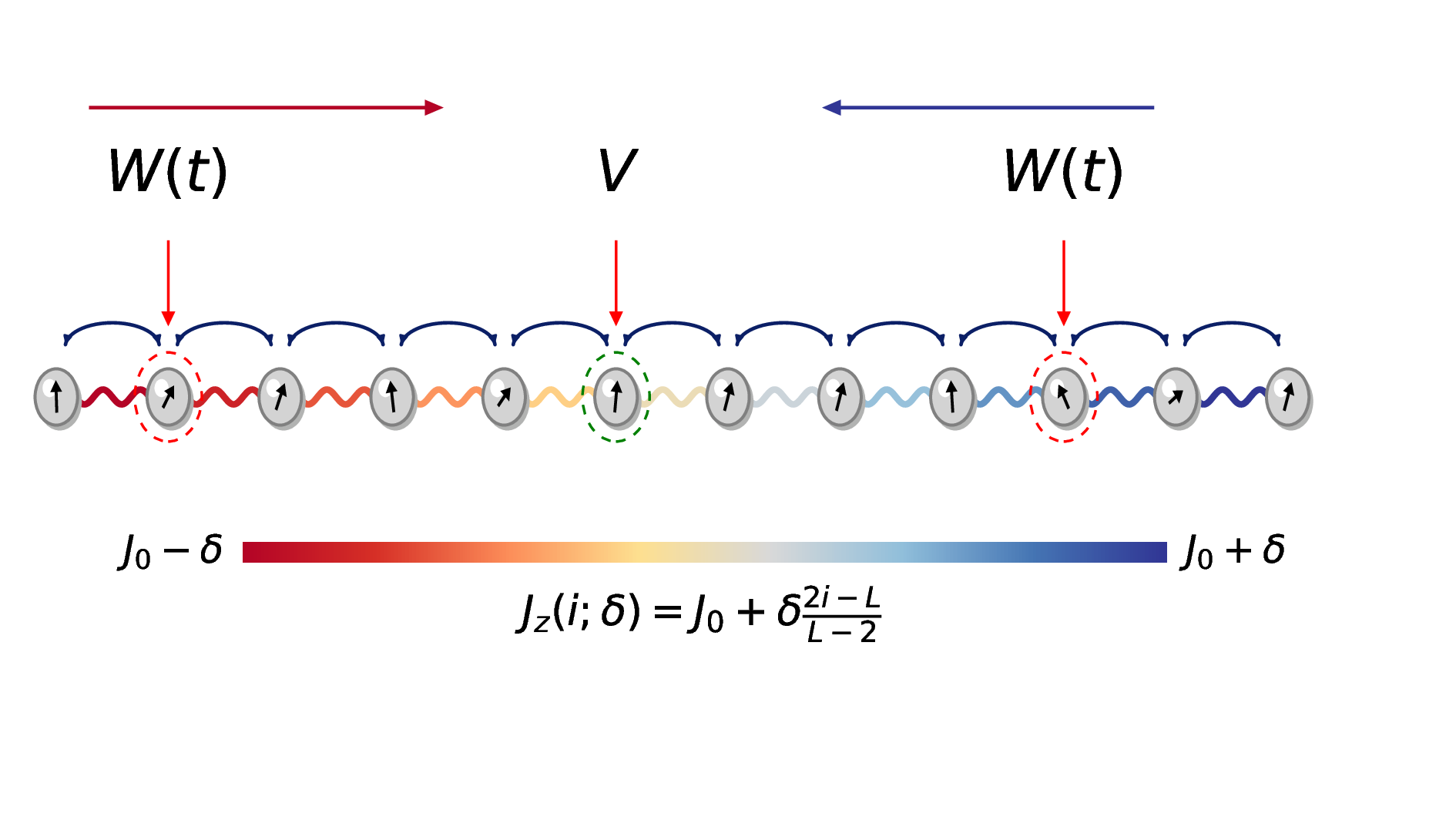}
		\vspace{-1cm}
		\caption{Schematic illustration of the asymmetric operator-propagation setup. A one-dimensional spin chain of length $L$ with open boundary conditions is considered. The nearest-neighbor Ising interaction varies spatially according to $J_z(i;\delta)=J_0+\delta\frac{2i-L}{L-2}$, whereas the hopping amplitude is uniform. Probe operators are initialized near the left or right boundary, and their dynamics are monitored through out-of-time-ordered correlators (OTOCs) with single-spin and bond operators located near the center of the chain. The interaction gradient breaks the left--right symmetry of operator propagation, leading to suppressed information scrambling for operators initialized on the strongly interacting (right) side compared with those on the weakly interacting (left) side.}
		\label{fig:operator_spreading}
	\end{figure}
	
	This article aims to investigate information scrambling in an inhomogeneous XXZ spin chain with spatially varying interactions, following the model introduced in Ref.~\cite{hamiltonian}. The central objective is to understand how the spatial inhomogeneity of the Hamiltonian manifests itself in the OTOC dynamics. We demonstrate that spatially varying interactions can give rise to an asymmetry in the dynamics of OTOCs and in their long-time finite-size saturation values, with OTOC operators located at different positions exhibiting distinct dynamical behavior. We further investigate the dynamics of OTOCs for a stepwise interaction profile and numerically show that, despite the presence of spatial inhomogeneity elsewhere in the system, the OTOCs exhibit nearly identical dynamics when the interaction strengths at the respective probe locations
	are the same. In contrast, when a sufficiently large difference in interaction strength is introduced between the two portions of the chain, the OTOC associated with the weakly interacting region exhibits faster scrambling than that associated with the strongly interacting region.
	These results provide a direct connection between the spatial interaction profile and the dynamics of OTOCs. 
	
	To understand the finite-size saturation values and relaxation dynamics, we interpret our results using the operator--Hamiltonian overlap framework (see, e.g., Refs.~\cite{Balachandran2021,Balachandran2023}). We show that when there is a nonvanishing overlap between the OTOC operators and the Hamiltonian, the OTOC exhibits a spatially dependent finite-size long-time saturation value and slow algebraic relaxation. In contrast, when the overlap vanishes, the conventional operator--Hamiltonian overlap picture predicts a vanishing long-time saturation value and relaxation faster than algebraic. However, our numerical results show that, for finite inhomogeneity the relaxation can remain slow even when the overlap vanishes, including in the chaotic regime. To understand the asymmetry in the finite-size, long-time saturation values of the OTOCs, we derive an analytical expression consistent with our numerical results. 
	To probe asymmetric information scrambling, we fix the static operator at the center of the chain, $i=L/2$, and place the time-evolving operator symmetrically near the left and right boundaries, at $i=2$ and $i=L-2$, respectively, as illustrated in Fig.~\ref{fig:operator_spreading}. To demonstrate the generality of our findings, we also present corresponding results for another interacting spin-chain model in Appendix~\ref{app:itfim_otoc}. We further extend our analysis to periodically driven inhomogeneous spin chains and show that the asymmetry in information scrambling persists in the driven systems (see Appendix~\ref{app:floquet_system}).
	
	The paper is structured as follows. In Sec.~\ref{sec:models}, we introduce the inhomogeneous XXZ Hamiltonian. Section~\ref{sec:level_stats} presents the spectral statistics and the crossover from integrability to quantum chaos. Section~\ref{sec:otoc} is devoted to the asymmetric information scrambling and relaxation dynamics revealed by the OTOCs. Finally, Sec.~\ref{sec:summary} summarizes our findings and presents the conclusions.
	
	\section{Inhomogeneous XXZ Chain}
	\label{sec:models}	
	We consider an inhomogeneous spin-$1/2$ XXZ chain with open boundary conditions. The Hamiltonian is given by
	\begin{equation}
		H(\delta)=	\sum_{i=1}^{L-1}\left[J_{xy}\left(S_i^xS_{i+1}^x+S_i^yS_{i+1}^y\right)
		+J_z(i;\delta)S_i^zS_{i+1}^z\right],
		\label{eq:H_total}
	\end{equation}
	where $S_i^{x,y,z}$ denote Pauli operators, with the individual spin operators having eigenvalues $\pm1$ (setting $\hbar=1$), and the spatially varying $z$--$z$ interaction strength is
	\begin{equation}
		J_z(i;\delta)=J_0+\delta\frac{2i-L}{L-2}.
		\label{eq:Jz_profile}
	\end{equation}
	Here, $\delta$ is the slope of the interaction gradient. Throughout this work, we consider hopping strength $J_{xy}=1$ and $J_0=1$.
	For $\delta=0$, the Hamiltonian reduces to the homogeneous integrable spin chain. As the inhomogeneity strength increases, the system exhibits a crossover from integrable behavior to quantum chaos and eventually toward a nonergodic regime for $\delta\gg1$.  
	The Hamiltonian conserves the total magnetization, $S^z_{\rm tot}=\sum_{i=1}^{L}S_i^z$, and all calculations are performed in the largest Hilbert-space sector with $S^z_{\rm tot}=0$.  The Hamiltonian also preserves the global spin-flip symmetry generated by $P^x=\prod_{i=1}^{L}S_i^x$. 
	To lift the degeneracies associated with the spin-flip symmetry in the \(S^z=0\) sector, we further resolve the spectrum into the even- and odd-parity sectors of the spin-flip symmetry. The dimension of the \(S^z=0\) Hilbert space is	$\frac{L!}{[(L/2)!]^2},$ which is reduced by a factor of two upon resolving the spin-flip symmetry, giving a dimension of 
	$\frac{L!}{2[(L/2)!]^2}$ for each parity sector.

	\section{Spectral Statistics and Finite-Size Scaling}
	\label{sec:level_stats}
	
	\begin{figure}[t]
		\centering
		\includegraphics[width=0.48\textwidth]{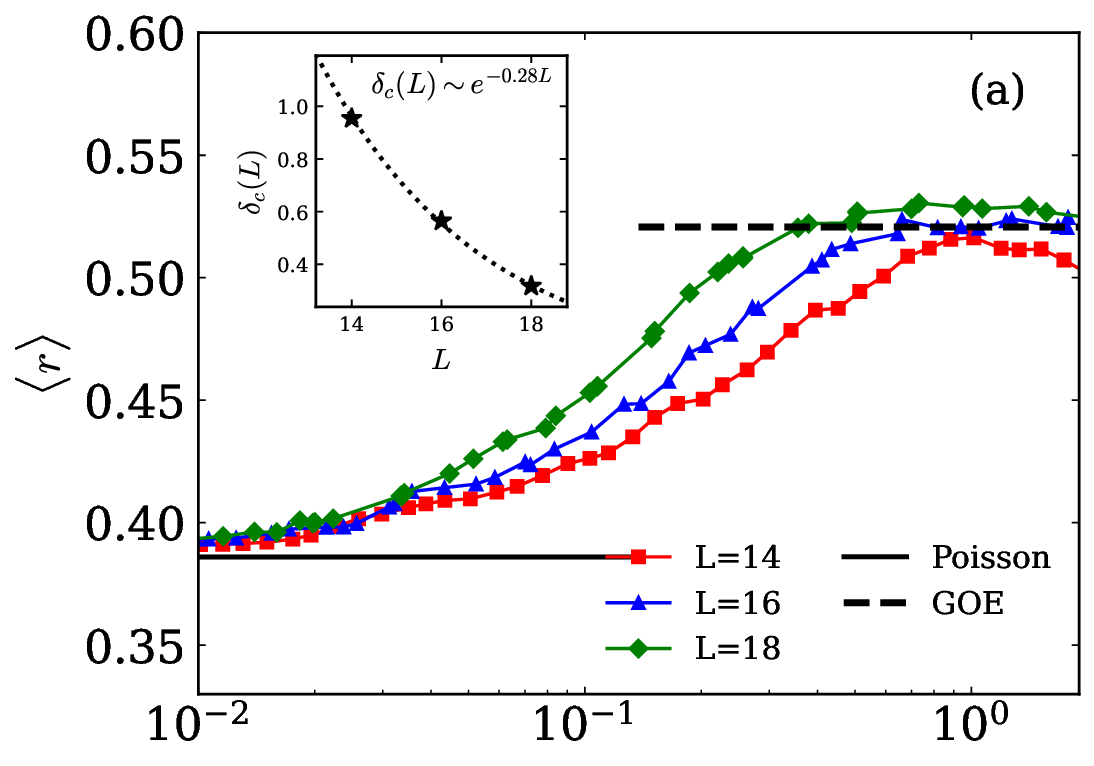}
		\includegraphics[width=0.48\textwidth]{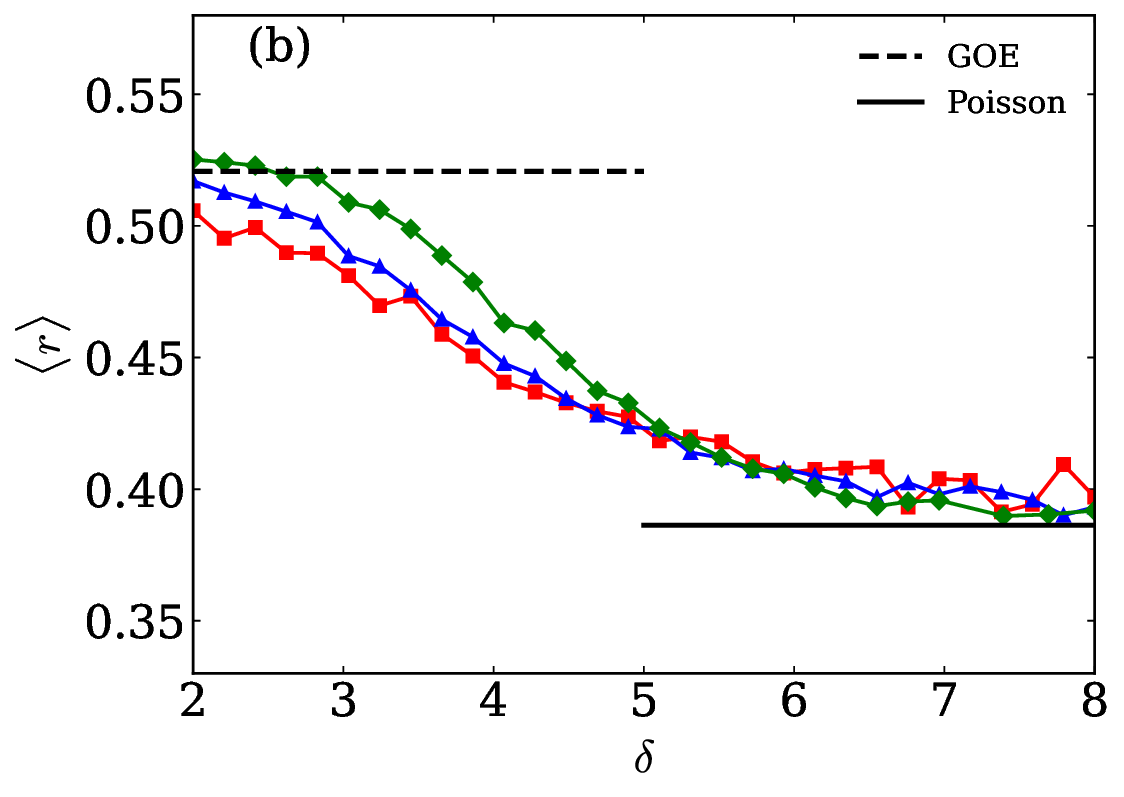}
		\caption{Evolution of the average consecutive level-spacing ratio $\langle r\rangle$ as a function of interaction-gradient strength $\delta$ for system sizes $L=14$, $16$, and $18$. The black solid and dashed horizontal lines denote the Poisson value, $\langle r\rangle\approx0.38$, and the GOE value, $\langle r\rangle\approx0.521$, respectively. (a) Starting from the homogeneous XXZ chain ($\delta=0$), $\langle r\rangle$ increases from the Poisson value toward the GOE prediction, indicating the crossover from integrability to quantum chaos with increasing interaction gradient. The inset shows the finite-size scaling of the critical value $\delta_c(L)$ corresponding to the onset of GOE statistics. (b) For larger interaction gradients, $\langle r\rangle$ gradually shifts from the GOE value toward the Poisson limit, indicating a crossover from the chaotic regime toward a regime with suppressed level repulsion induced by strong interaction inhomogeneity.}
		
		\label{fig:level_statistics}
	\end{figure}
	
	We begin by analyzing the spectral statistics through the average ratio of consecutive levels
	of the inhomogeneous XXZ chain defined in Eq.~(\ref{eq:H_total}). In the homogeneous limit ($\delta=0$), the model reduces to the integrable spin chain, whose energy-level statistics follow the Poisson distribution, reflecting the absence of level repulsion. As the interaction gradient increases, integrability is progressively broken, giving rise to level repulsion and driving the spectrum toward the Gaussian orthogonal ensemble (GOE), characteristic of quantum chaotic systems.
	
	To characterize the spectral correlations, we compute the average ratio of consecutive level
	spacings \cite{spacing_rat,spacing_rat1}, defined as,
	\begin{equation}
		r_n	= \frac{\min(s_n,s_{n+1})}{\max(s_n,s_{n+1})},
		\qquad
		s_n=E_{n+1}-E_n,
		\label{eq:level_ratio}
	\end{equation}
	where $E_n$ are the ordered eigenvalues of the Hamiltonian. The mean level spacing ratio $\langle r \rangle$ provides a convenient measure of spectral correlations without requiring unfolding of the spectrum. For integrable systems with uncorrelated eigenvalues, the spacing-ratio distribution follows the Poisson statistics, $P_{\mathrm{Poi}}(r)=\frac{2}{(1+r)^2}$, with $\langle r\rangle\approx0.386$. In contrast, for quantum-chaotic systems described by GOE, the distribution is $P_{\mathrm{GOE}}(r)=\frac{27}{4}\frac{r+r^2}{(1+r+r^2)^{5/2}}$, with $\langle r\rangle\approx0.5307$ \cite{spacing_rat,spacing_rat1}.
	
	Figure~\ref{fig:level_statistics} shows the evolution of the average ratio of consecutive level spacings \ $\langle r\rangle$ with increasing $\delta$ for different system sizes. For small $\delta$, the $\langle r\rangle$ remain close to the Poisson limit. As the inhomogeneity increases, $\langle r\rangle$ approaches the GOE prediction, indicating the onset of quantum chaos induced by the deterministic interaction gradient. On further increasing the interaction gradient ($\delta\gg1$), the $z$--$z$ coupling dominates the Hamiltonian, while the hopping term becomes too weak to mix the many-body basis states significantly. Consequently, the eigenstates remain weakly delocalized in the $S_i^z$ basis, leading to a reemergence of Poissonian level statistics.

	\section{Relaxation dynamics and finite-size saturation value of OTOCs}
	\label{sec:otoc}
	
	The OTOC provides a measure of the growth of operator noncommutativity and hence serves as a diagnostic of information scrambling in quantum systems. For two operators $W$ and $V$, we define 
	The infinite-temperature OTOC (with $\langle \cdots \rangle = \mathrm{Tr}(\cdots)/\mathcal{D}$) as 
	\begin{equation}
		\mathcal{C}_{WV}(t) = -\frac{1} {2\mathcal{D}} \mathrm{Tr}\left([W(t),V]^2\right),
		\label{eq:otoc_def}
	\end{equation}
	where $W(t)=e^{iHt}We^{-iHt}$ and $\mathcal{D}$ is the Hilbert space dimension, $1/2$ factor is taken for normalization purpose. For Hermitian unitary operators, the squared commutator simplifies to
	\begin{align}
		\mathcal{C}_{WV}(t)	&=1	-\mathrm{Re}\,
		\mathcal{F}_{WV}(t),
		\nonumber\\
		\mathcal{F}_{WV}(t)
		&=\frac{1}{\mathcal{D}} \mathrm{Tr}
		\left(
		W(t)VW(t)V
		\right).
		\label{eq:four_point}
	\end{align}
	where $\mathcal{F}_{WV}(t)$ is the four-point out-of-time-ordered correlator. This quantity directly encodes the dynamical correlations between initially commuting operators and determines the relaxation behavior of the OTOC.
	
	\subsection{Infinite-Time Value of OTOC }
	
	To gain analytical insight into the infinite-time limit of OTOC, we express $\mathcal{F}_{WV}(t)$ [see Eq.~(\ref{eq:four_point})] in the complete basis of the Hamiltonian energy eigenstates $\{|\alpha\rangle\}$ with eigenvalues $E_\alpha$ satisfying $H|\alpha\rangle = E_\alpha |\alpha\rangle$,
	as
	\begin{equation}
		\mathcal{F}_{WV}(t)
		=
		\frac{1}{\mathcal{D}}
		\sum_{\alpha\beta\gamma\delta}
		e^{i(E_\alpha-E_\beta+E_\gamma-E_\delta)t}
		W_{\alpha\beta}
		V_{\beta\gamma}
		W_{\gamma\delta}
		V_{\delta\alpha},
		\label{eq:spectral}
	\end{equation}
	where $W_{\alpha\beta}=\langle\alpha|W|\beta\rangle$ and $V_{\alpha\beta}=\langle\alpha|V|\beta\rangle$. The long-time behavior is determined by the infinite-time average of Eq.~(\ref{eq:spectral}). Under the nondegenerate energy-gap condition
	$E_\alpha+E_\gamma=E_\delta+E_\beta
	\quad \Rightarrow \quad
	(\alpha=\beta,\gamma=\delta)
	\ \text{or}\
	(\alpha=\delta,\beta=\gamma)$,
	all oscillatory contributions dephase at long times, leaving only the diagonal and paired off-diagonal terms. The infinite-time average therefore becomes	
	\begin{align}
		\mathcal{F}_{WV}(\infty) &= \frac{1}{\mathcal{D}} \sum_{\alpha\neq\beta} \Big(
		W_{\alpha\alpha}
		W_{\beta\beta}
		|V_{\alpha\beta}|^2
		+
		V_{\alpha\alpha}
		V_{\beta\beta}
		|W_{\alpha\beta}|^2
		\Big)
		\nonumber\\
		&\quad+
		\frac{1}{\mathcal{D}}
		\sum_{\alpha}
		W_{\alpha\alpha}^2
		V_{\alpha\alpha}^2.
		\label{eq:infinite_avg}
	\end{align}	
	According to eigenstate thermalization hypothesis (ETH), the diagonal matrix elements are smooth functions of energy,
	$W_{\alpha\alpha}=f_W(E_\alpha)$ and $V_{\alpha\alpha}=f_V(E_\alpha)$. Consequently, for pairs of eigenstates connected by low-frequency matrix elements, for which $|E_\alpha-E_\beta|\ll 1$, the eigenstate-to-eigenstate fluctuations of the diagonal matrix elements are exponentially suppressed
	with system size in chaotic systems \cite{21,23,25,mishra2026}. Thus, within this low-frequency regime, eigenstates with sufficiently close energies have nearly identical local expectation values, allowing us to approximate
	$W_{\beta\beta}\approx W_{\alpha\alpha}$ and $V_{\beta\beta}\approx V_{\alpha\alpha}$.
	Using the identities
	\begin{equation}
		\sum_{\beta} W_{\alpha\beta} V_{\beta\alpha} = (WV)_{\alpha\alpha},
		\qquad
		\sum_{\beta} V_{\alpha\beta} W_{\beta\alpha} = (VW)_{\alpha\alpha},
	\end{equation}
	the off-diagonal sums can be written as
	\begin{align}
		\sum_{\beta\neq\alpha}
		W_{\beta\beta}
		V_{\alpha\beta}
		V_{\beta\alpha}
		&\approx
		W_{\alpha\alpha}
		\sum_{\beta\neq\alpha}
		V_{\alpha\beta}
		V_{\beta\alpha}
		\nonumber\\
		&=
		W_{\alpha\alpha}
		\Big[
		(V^2)_{\alpha\alpha}
		-
		V_{\alpha\alpha}^2
		\Big],
		\\
		\sum_{\beta\neq\alpha}
		V_{\beta\beta}
		W_{\alpha\beta}
		W_{\beta\alpha}
		&\approx
		V_{\alpha\alpha}
		\sum_{\beta\neq\alpha}
		W_{\alpha\beta}
		W_{\beta\alpha}
		\nonumber\\
		&=
		V_{\alpha\alpha}
		\Big[
		(W^2)_{\alpha\alpha}
		-
		W_{\alpha\alpha}^2
		\Big].
	\end{align}
	For unitary operators, $W^2 = V^2 = I$, so $(W^2)_{\alpha\alpha} = (V^2)_{\alpha\alpha} = 1$.
	Substituting these expressions into Eq.~(\ref{eq:infinite_avg}) yields
	\begin{align}
		\mathcal{F}_{WV}(\infty)
		&\approx
		\frac{1}{\mathcal{D}}
		\sum_{\alpha}
		\Big[
		W_{\alpha\alpha}^{2}
		\left(1-V_{\alpha\alpha}^{2}\right)
		+
		V_{\alpha\alpha}^{2}
		\left(1-W_{\alpha\alpha}^{2}\right)
		\nonumber\\
		&\qquad\qquad
		+
		W_{\alpha\alpha}^{2}V_{\alpha\alpha}^{2}
		\Big]
		\nonumber\\
		&\approx
		\frac{1}{\mathcal{D}}
		\sum_{\alpha}
		\left(
		W_{\alpha\alpha}^{2}
		+
		V_{\alpha\alpha}^{2}
		-
		W_{\alpha\alpha}^{2}V_{\alpha\alpha}^{2}
		\right).
		\label{eq:FWV_inf}
	\end{align}
	Equation~(\ref{eq:FWV_inf})  should therefore be understood as an ETH-based leading-order	approximation to the infinite-time OTOC. Corrections arise from energy-separated eigenstate pairs for which the smooth-energy approximation of the diagonal matrix elements is no longer valid.
	Equation~(\ref{eq:FWV_inf}) establishes a direct connection between the OTOC saturation value and the diagonal matrix elements of the operators in the energy eigenbasis. 
	Consequently, operators with larger diagonal weights exhibit larger finite-size saturation values, whereas vanishing diagonal matrix elements correspond to saturation values close to zero.
	
	\subsection{Influence of Operator--Hamiltonian Overlap on OTOC Dynamics}

	\begin{figure}[t]
		\centering
		\includegraphics[width=0.48\textwidth]{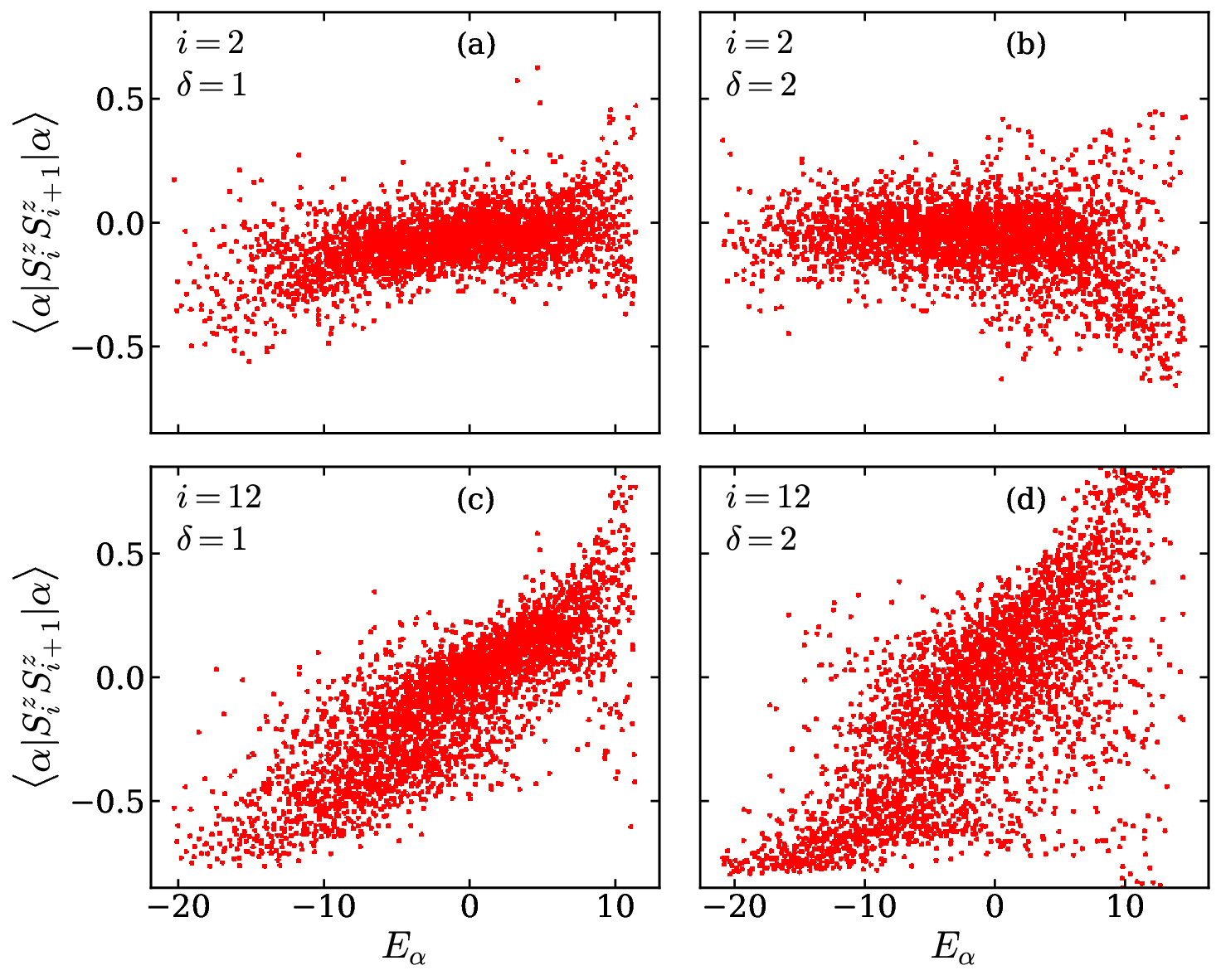}
		\includegraphics[width=0.48\textwidth]{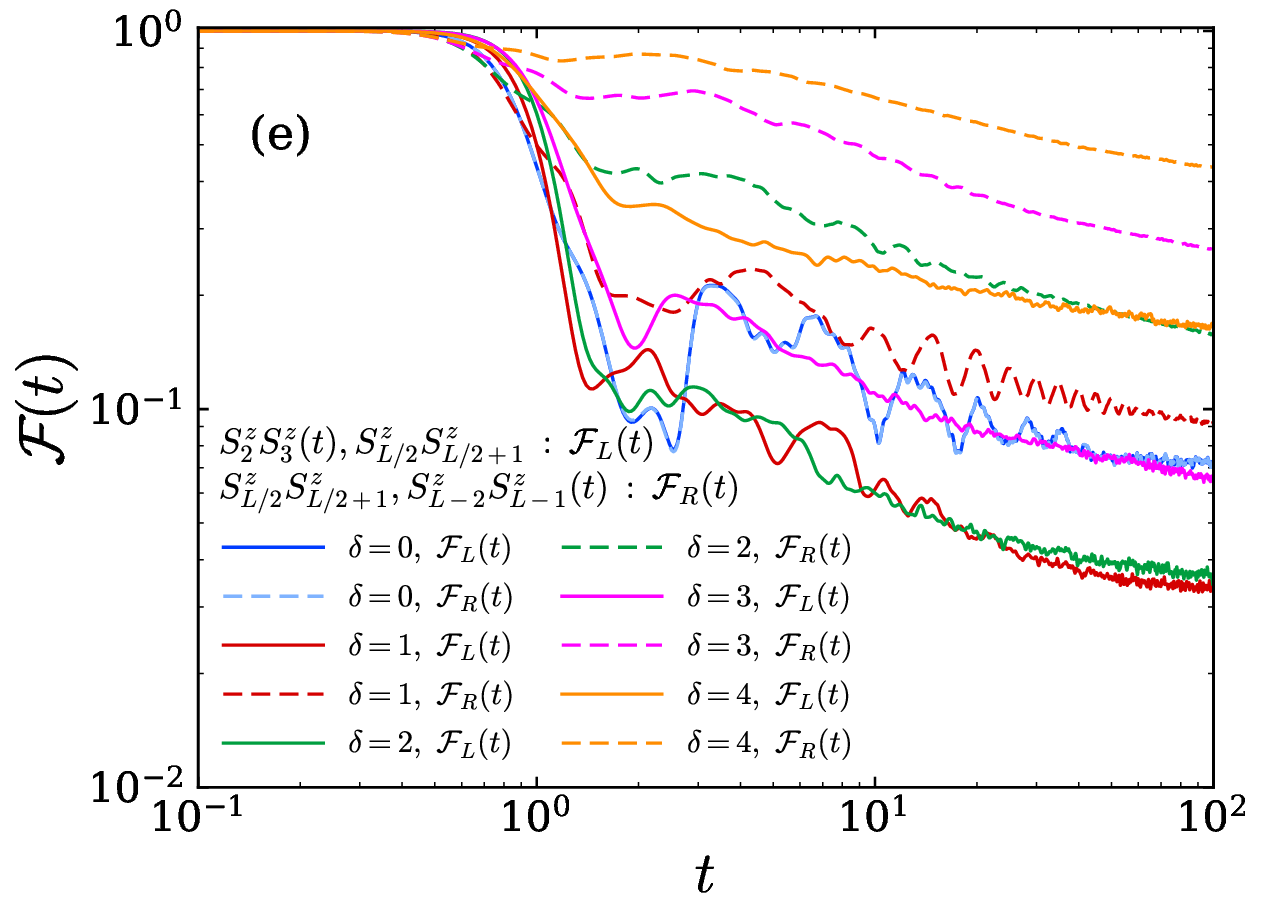}
		\caption{(a),(b) Diagonal matrix elements $\langle\alpha|S_i^zS_{i+1}^z|\alpha\rangle$ in the energy eigenbasis of the Hamiltonian defined in Eq.~\ref{eq:H_total} for $i=2$, and (c),(d) for $i=L-2$, plotted as functions of the eigenenergy $E_\alpha$. Panels (a),(c) and (b),(d) correspond to $\delta=1$ and $2$, respectively. (e) OTOC dynamics for the left and right probes, $\mathcal{F}_L(t)$ and $\mathcal{F}_R(t)$, for interaction gradients $\delta=0$, $1$, $2$, $3$, and $4$. Within the intermediate-time window, $0.7\lesssim t\lesssim3$, the OTOC associated with the weakly interacting side decays more rapidly than that of the strongly interacting side. Beyond this intermediate-time regime, the OTOCs exhibit algebraic relaxation toward their respective saturation values. The smaller weight of the diagonal matrix elements on the weakly interacting side (left) is associated with a lower OTOC saturation value. Results are shown for system size $L=14$.}
		\label{fig:diag_eth_and_otoc_Sz_NN}
	\end{figure}
	\begin{figure}[t]
		\centering
		\includegraphics[width=0.48\textwidth]{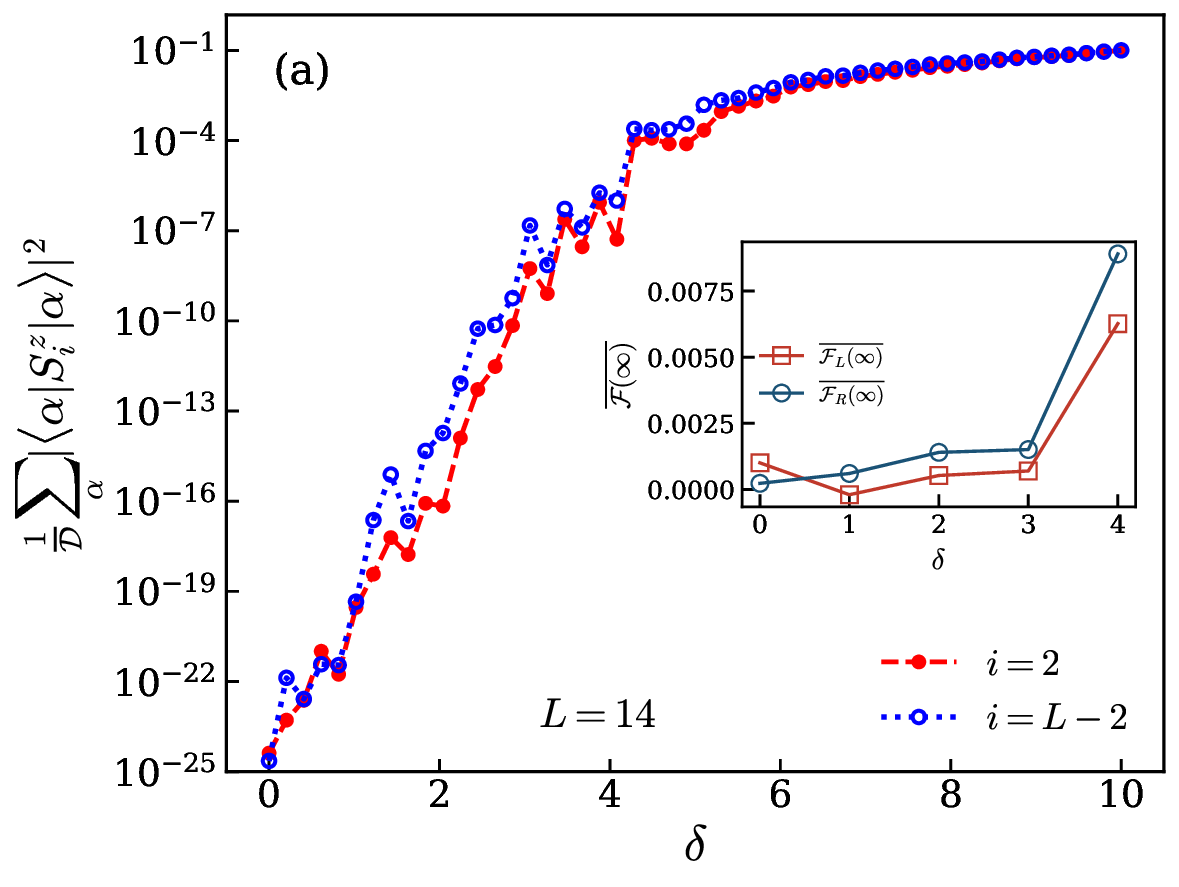}
		\includegraphics[width=0.48\textwidth]{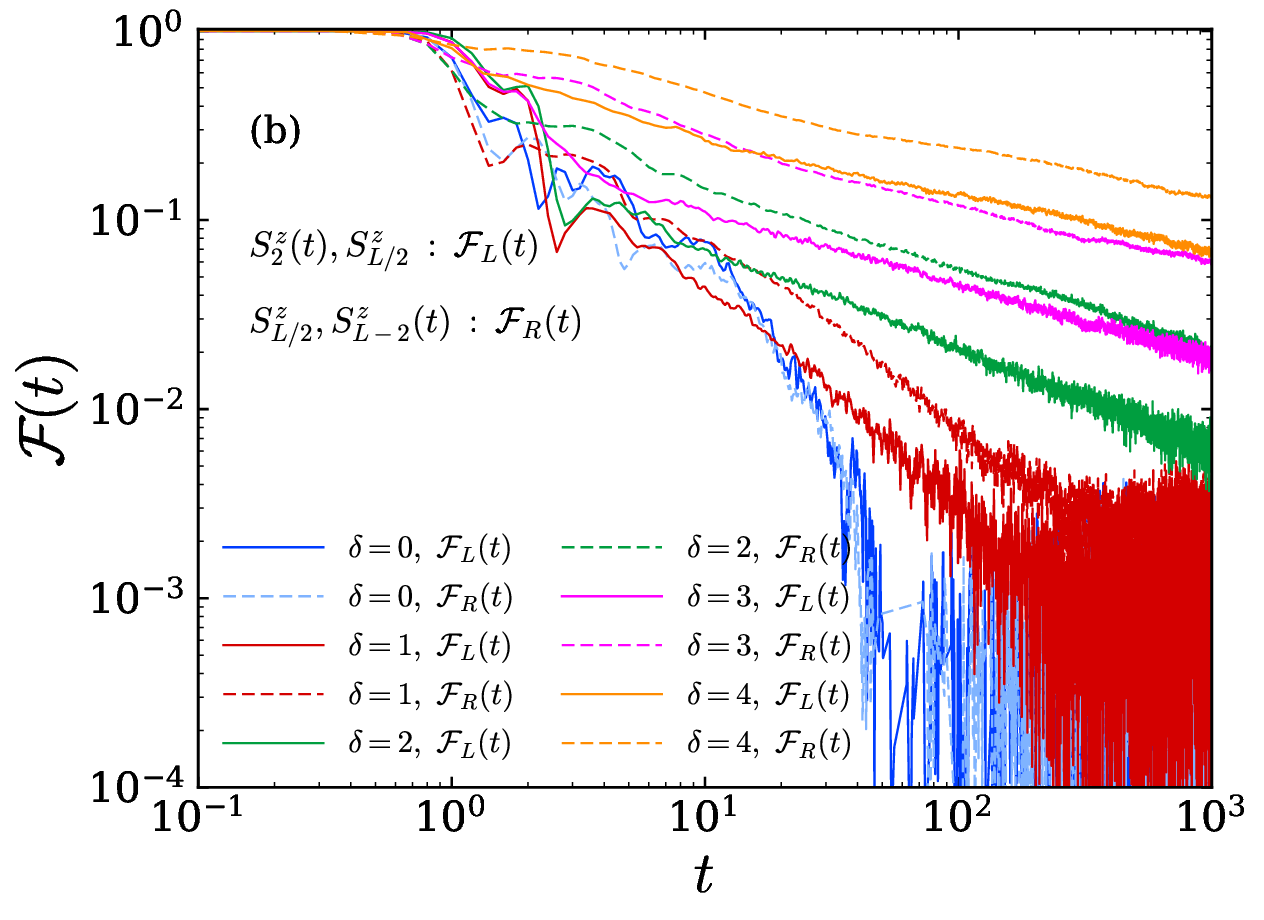}
		\caption{(a) Averaged squared diagonal matrix elements as a function of the interaction-gradient strength $\delta$ for the local spin operator $S_i^z$ at sites $i=2$ and $i=L-2$. The averaged squared diagonal matrix elements remain vanishingly small, with $\langle\alpha|S_i^z|\alpha\rangle=0$, as enforced by Eq.~(\ref{eq:parity_zero}). (b) OTOC dynamics for the left and right probes, $\mathcal{F}_L(t)$ and $\mathcal{F}_R(t)$, for $\delta=0$, $1$, $2$, $3$, and $4$. The inset of panel (a) shows the corresponding long-time saturation values of the OTOCs, obtained by averaging over twenty time points around $t=\mathcal{O}(10^9)$. All results are shown for system size $L=14$.}
		\label{fig:OTOC_local_Sz_h0}
	\end{figure} 
	
	The relaxation of OTOCs is intimately connected to the overlap between the OTOC operators and the Hamiltonian~\cite{analytical1}. For Hamiltonians with local interactions, the Lieb--Robinson bound implies that the OTOC dynamics in the thermodynamic limit can be understood from the infinite-time behavior of a finite system of effective size $L_{\rm LR}\sim v_{\rm LR}t$. Thus, the OTOC in the thermodynamic limit can be related to the long-time behavior of a finite system of size
	$L_{\rm LR}$. Once the system is fully scrambled, one further has $F_{WV}^{L=\infty}(t)\approx F_{WV}^{L_{\rm LR}}(\infty)$, so that the relaxation dynamics are determined by the finite-size scaling of the infinite-time saturation value (see, e.g., Refs.~\cite{Balachandran2021,Balachandran2023,Balachandran_entropy}). Equation~(\ref{eq:FWV_inf}) shows that this saturation value is governed by the diagonal matrix elements of OTOC operators in the energy eigenbasis. When these operators possess a finite overlap with the Hamiltonian, the infinite-time value scales as $F_{WV}(\infty)\propto L^{-\eta}$, leading directly to the algebraic relaxation law $F_{WV}(t)\propto t^{-\eta}$~\cite{Balachandran2023,Balachandran_entropy}.
	
	\begin{figure*}[t]
		\centering
		\includegraphics[width=\textwidth]{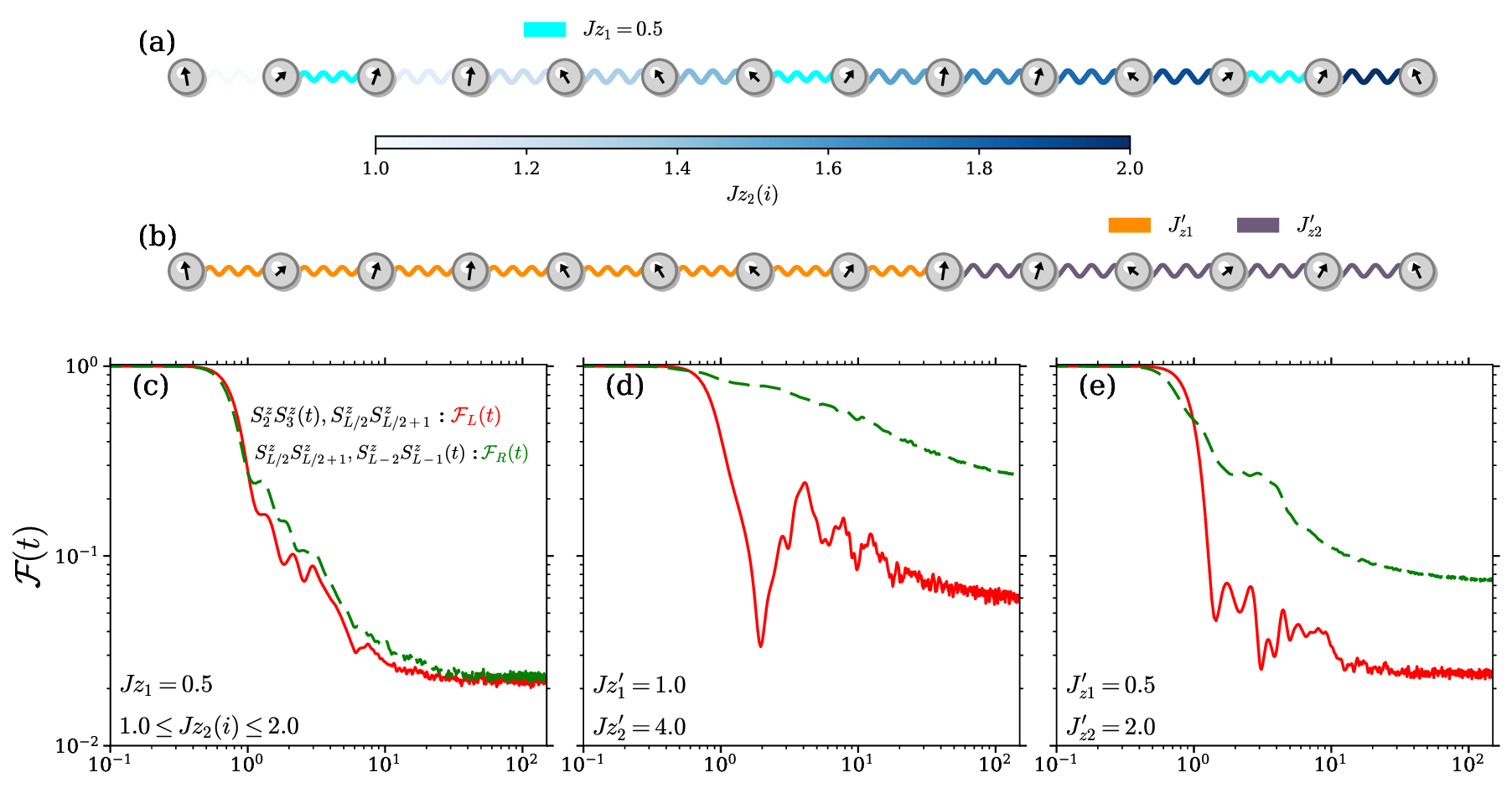}
		\includegraphics[width=\textwidth]{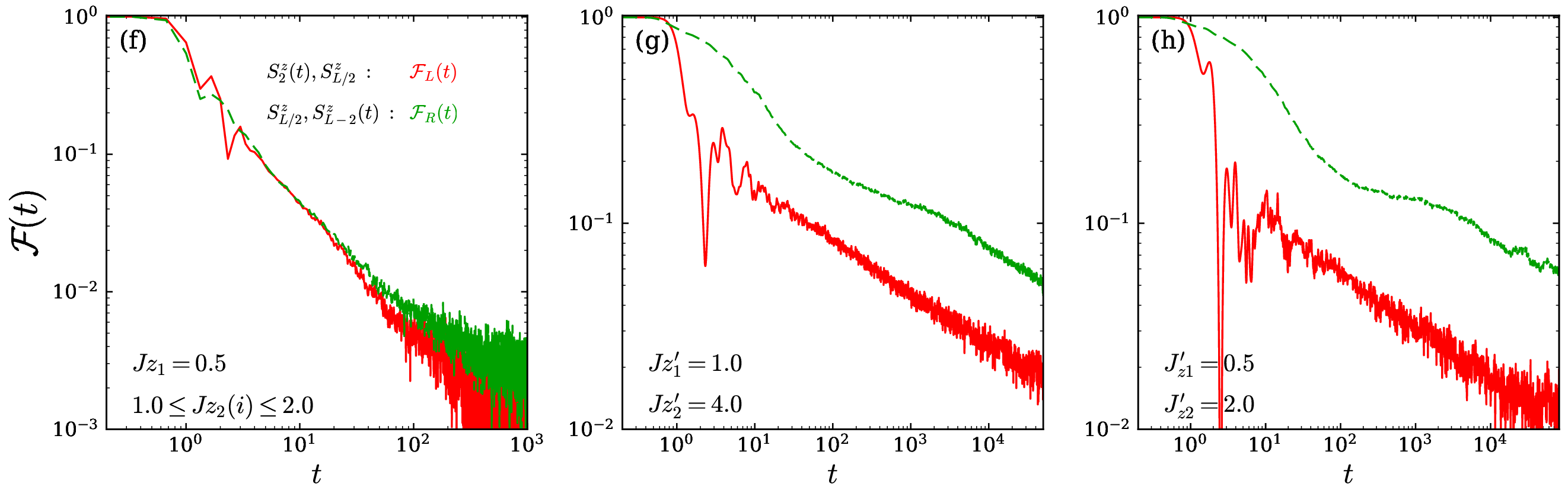}
		\caption{OTOC dynamics for a stepwise inhomogeneous XXZ chain as represented in Eqs. \ref{eq:Jz_smooth} and \ref{eq:Jz_piecewise}. In the setup (a), the nearest-neighbor interaction strength varies as, with $J_{z1}=0.5$ and $1.0\leq J_{z2}(i)\leq2.0$. (b) Piecewise interaction profile with two distinct couplings, $J'_{z1}$ and $J'_{z2}$, assigned to the left and right portions of the chain, respectively. OTOCs corresponding to the left and right sides are represented by $\mathcal{F}_{L}(t)$ (solid red lines) and $\mathcal{F}_{R}(t)$ (dashed green lines), respectively. In (c) and (f), the OTOC operators are located in equivalent local interaction regions, while the interaction strengths at the remaining sites vary according to the smooth profile shown in (a), resulting in nearly identical OTOC dynamics. In (d), (e), (g), and (h), the interaction strengths in the two portions of the chain differ, as shown in (b), leading to distinct dynamical behavior of the OTOCs. In particular, the OTOC associated with the strongly interacting region exhibits suppressed scrambling compared with that in the weakly interacting region.}
		\label{fig:stepwise_interaction}
	\end{figure*}
	\begin{figure}[]
		\centering
		\includegraphics[width=0.48\textwidth]{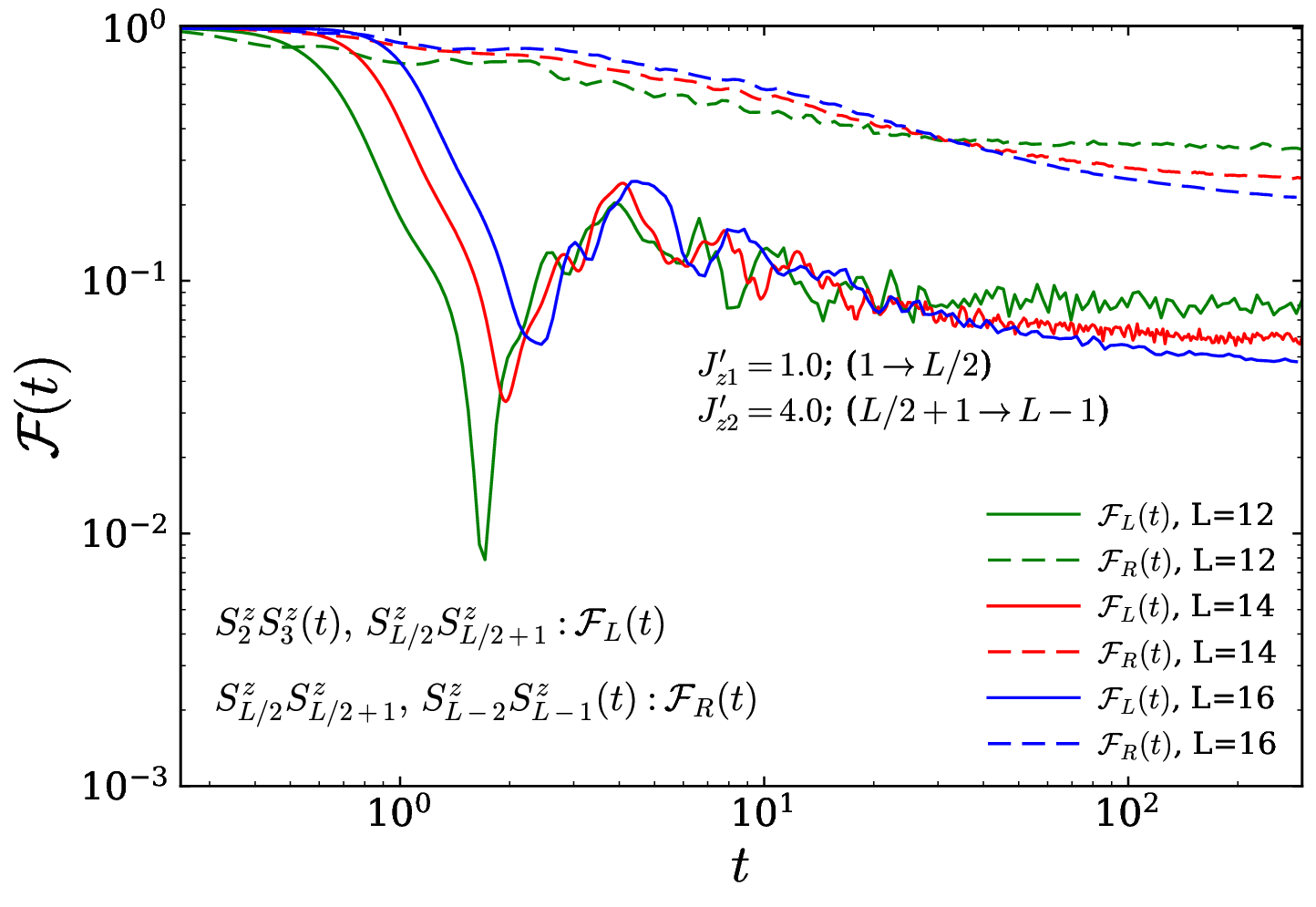}
		\caption{OTOC dynamics for the bond operator $S_i^zS_{i+1}^z$ in a piecewise inhomogeneous XXZ chain. Results are shown for system sizes $L=12$, $14$, and $16$. The strongly interacting side exhibits suppressed scrambling, with slower relaxation of the corresponding OTOC than on the weakly interacting side. Despite the decrease of the finite-size saturation value with increasing system size, the left--right asymmetry in the OTOC dynamics persists for larger chains.}
		\label{fig:OTOC_SzNN_L}
	\end{figure}
	To study the OTOC dynamics, we consider two representative local observables: the bond operator $A_i = S_i^z S_{i+1}^z$ and the local spin operator $B_i = S_i^z$. In the $S^z_{\rm tot}=0$ sector, the Hamiltonian commutes with the spin-flip operator $P^x = \prod_i S_i^x$, which anticommutes with $S_i^z$ and commutes with $S_i^z S_{i+1}^z$. For even $L$, the energy eigenstates can be chosen to have definite spin-flip parity,
	$P^x|\alpha\rangle = \pm|\alpha\rangle$. Consequently,
	\begin{equation}
		\langle \alpha | S_i^z | \alpha \rangle
		= \langle \alpha | P^x S_i^z P^x | \alpha \rangle
		= -\langle \alpha | S_i^z | \alpha \rangle = 0.
		\label{eq:parity_zero}
	\end{equation}
	Thus, the diagonal matrix elements of $S_i^z$ vanish as a consequence of the global spin-flip symmetry. In contrast, the bond operator $S_i^z S_{i+1}^z$ is not constrained to vanish by this symmetry. The site dependence of the diagonal matrix elements of the bond operator as a function of the local interaction strength $J_z(i;\delta)$ is discussed in Appendix~\ref{app:diagonal_NN}.

To compare the scrambling dynamics on the left and right sides of the chain on an equal footing, we choose the probe operators with the same spatial separation $(i-j=d)$. For the bond operator $S_i^zS_{i+1}^z$, the left and right probes are placed symmetrically with respect to the center of the chain, making them equivalent in the homogeneous limit ($\delta=0$) and resulting in identical OTOC dynamics. In contrast, the local spin operators $S_i^z$ are located at sites $i=2$ and $i=L-2$, which are not exactly equivalent under open boundary conditions. Consequently, even for $\delta=0$, the two probes experience slightly different boundary effects, leading to small differences in their OTOC dynamics. However, for onsite operators in the homogeneous case, similar left--right OTOC dynamics can be obtained for odd system sizes.

  We first consider the bond operator $S_i^zS_{i+1}^z$, for which the operator--Hamiltonian overlap is finite. Consequently, the corresponding diagonal matrix elements remain finite in the energy eigenbasis. Figure~\ref{fig:diag_eth_and_otoc_Sz_NN}(a)--(d) shows that the diagonal matrix elements exhibit a pronounced dependence on both the site and the interaction gradient. In particular, the total diagonal weight, $\sum_{\alpha}W_{\alpha\alpha}^2$, is significantly larger on the strongly interacting (right) side of the chain than on the weakly interacting (left) side. According to Eq.~(\ref{eq:FWV_inf}), this directly results in a larger long-time saturation value of the OTOC.
  The corresponding OTOCs shown in Fig.~\ref{fig:diag_eth_and_otoc_Sz_NN}(e) exhibit the expected algebraic relaxation, consistent with the finite operator--Hamiltonian overlap. For the homogeneous chain ($\delta=0$), the left and right OTOCs show identical behavior. As the interaction gradient increases, a pronounced asymmetry emerges in the OTOC dynamics. 
  in the intermediate-time regime the OTOC associated with the weakly interacting side,  $\mathcal{F}_{L}(t)$, exhibits a steeper initial decay, indicating fast scrambling, and subsequently saturates at a lower value. In contrast, the OTOC on the strongly interacting side, $\mathcal{F}_{R}(t)$, decays more gradually, indicating suppressed scrambling, and saturates at a higher value.

   For the local spin operator $S_i^z$, Eq.~(\ref{eq:parity_zero}) enforces
  $\langle\alpha|S_i^z|\alpha\rangle=0$ for every eigenstate, within numerical precision [see Fig.\ref{fig:OTOC_local_Sz_h0} (a)]. Consequently, Eq.~(\ref{eq:FWV_inf}) predicts a vanishing long-time saturation value; however, the numerical results exhibit a small but nonzero saturation value due to finite-size effects.  The conventional operator--Hamiltonian overlap approach therefore predicts relaxation faster than algebraic (possibly exponential) rates. Surprisingly, the numerical results shown in Fig.~\ref{fig:OTOC_local_Sz_h0}(b) reveal that the relaxation toward the saturation becomes progressively slower as the interaction gradient $\delta$ increases, even in the quantum-chaotic regime, despite the absence of diagonal matrix elements. Moreover, although Eq.~(\ref{eq:FWV_inf}) predicts an approximately vanishing long-time saturation value, a pronounced left--right asymmetry persists in the OTOC dynamics, similar to that observed for the bond operator.

  We summarize this section as follows: the operator--Hamiltonian overlap framework does not fully account for the observed relaxation dynamics in the present inhomogeneous system. Our numerical results indicate that the relaxation becomes slower with increasing inhomogeneity. For local $S_i^z$ operators, symmetry enforces vanishing diagonal matrix elements, resulting in nearly identical long-time saturation values for the left and right probes [inset of Fig.~\ref{fig:OTOC_local_Sz_h0}(a)]. For operators with non-vanishing overlap with the Hamiltonian, the finite-size long-time saturation scales as $\mathcal{F}(\infty)\propto 1/L$ Fig.~\ref{fig:OTOC_longtime}(a)~\cite{Balachandran2021,Balachandran_entropy} and therefore vanishes in the thermodynamic limit. Nevertheless, the local interaction strength remains inequivalent as $L$ increases. Since operator growth is governed by the local Hamiltonian and constrained by the Lieb--Robinson bound~\cite{Lib_rob2}, differences in the local interaction profiles lead to distinct OTOC dynamics even when the long-time saturation values of the left and right probes coincide or vanish.
\subsection{OTOC Dynamics in a Stepwise Inhomogeneous Interaction}

To further distinguish the role of the local interaction from the mere presence of spatial inhomogeneity, we consider a piecewise interaction	profile in the $z$ direction, as illustrated in Fig.~\ref{fig:stepwise_interaction}. The Hamiltonian is given by
\begin{equation}
	H =\sum_{i=1}^{L-1}	\left[\left(S_i^xS_{i+1}^x+S_i^yS_{i+1}^y\right)+J_z(i)S_i^zS_{i+1}^z \right].
	\label{eq:piecewise_H}
\end{equation}
The spatial dependence of the interaction strength is implemented in two distinct ways, as illustrated in Fig.~\ref{fig:stepwise_interaction}(a) and Fig.~\ref{fig:stepwise_interaction}(b), respectively, and is given by

\begin{equation}
	J_z^{(a)}(i)=
	\begin{cases}
		J_{z1}(i)=0.5, & i=2,\;L/2,\;L-2,\\
		J_{z2}(i)=1+\dfrac{i-1}{L-2}, & \text{otherwise},
	\end{cases}
	\label{eq:Jz_smooth}
\end{equation}

\begin{equation}
	J_z^{(b)}(i)=
	\begin{cases}
		J_{z1}', & 1\leq i\leq L/2,\\
		J_{z2}', & L/2<i\leq L-1,
	\end{cases}
	\label{eq:Jz_piecewise}
\end{equation}
  In the setup represented in Fig.~\ref{fig:stepwise_interaction}(a), the interaction strength varies linearly while remaining constant at the locations of the OTOC operators. Interestingly, as shown in Figs.~\ref{fig:stepwise_interaction}(c) and (f), both probe operators experience the same local interaction strength, resulting in nearly identical dynamics and long-time saturation values of $\mathcal{F}_{L}(t)$ and $\mathcal{F}_{R}(t)$. The corresponding average level-spacing ratio of the Hamiltonian is $\langle r\rangle\approx0.52$.

  A qualitatively different behavior emerges when the two probe operators are embedded in regions with different interaction strengths, as shown in Figs.~\ref{fig:stepwise_interaction}(d), (e), (g), and (h). The OTOCs exhibit pronounced asymmetry in the scrambling dynamics, with the probe associated with the strongly interacting region exhibiting suppressed scrambling. For the case shown in Figs.~\ref{fig:stepwise_interaction}(d) and (g), the average level-spacing ratio is $\langle r\rangle\approx0.46$, indicating a departure from the ergodic regime, while for the case shown in Figs.~\ref{fig:stepwise_interaction}(e) and (h), it is $\langle r\rangle\approx0.53$. Thus, the observed asymmetry occurs within the nonintegrable regime and becomes more pronounced as the difference between the two interaction strengths increases.
 Finally, Fig.~\ref{fig:OTOC_SzNN_L} demonstrates the asymmetry in information scrambling for increasing system sizes.

\subsection{Results for the Long-Time Saturation of OTOCs}

\begin{figure*}[htbp]
	\centering	
	\includegraphics[width=\textwidth]{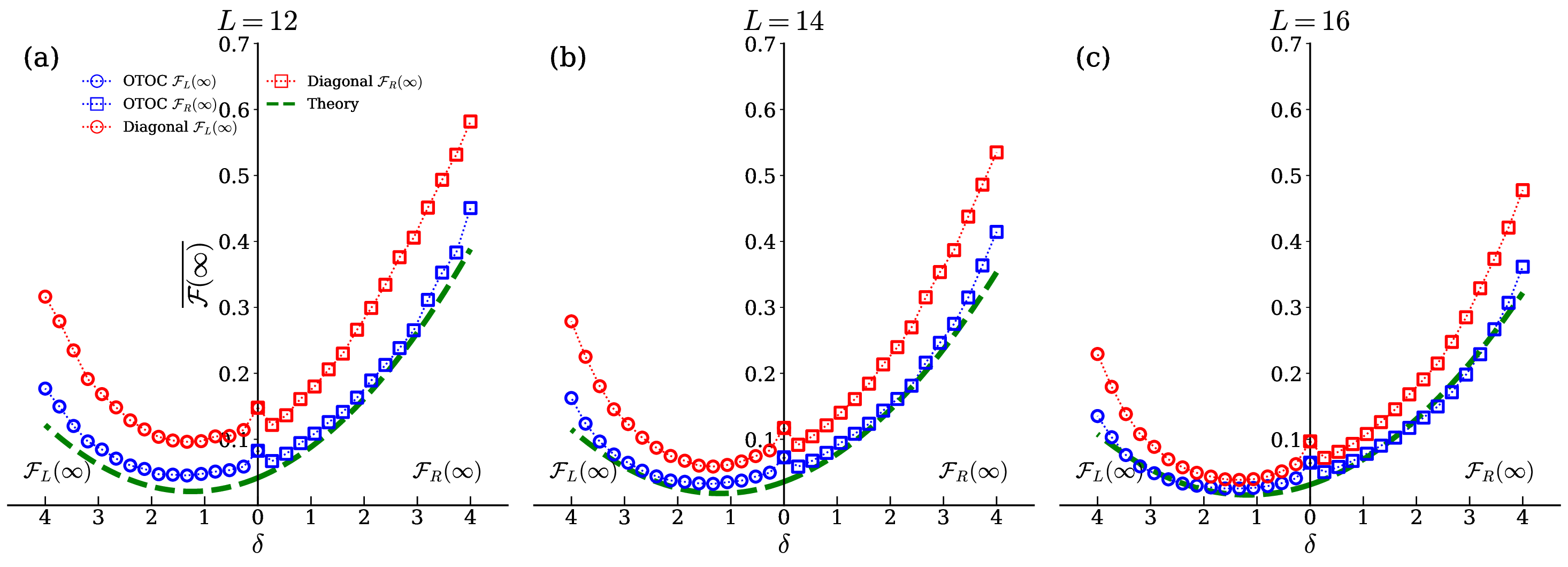}
	\caption{Finite-size saturation values of the OTOCs for the bond operator $S_i^zS_{i+1}^z$, $\mathcal{F}_L(\infty)$ and $\mathcal{F}_R(\infty)$, as functions of the interaction gradient $\delta$ for different system sizes. The open red symbols denote the OTOC saturation values predicted by the diagonal matrix elements using Eq.~(\ref{eq:FWV_inf}), while the open blue symbols represent the numerical long-time averages obtained by averaging over the twenty time points at $\mathcal{O}(10^9)$. The dashed green line represents the analytical prediction from Eq.~(\ref{eq:PRL_result}). The small deviations between the analytical prediction and the numerical results are due to finite-size effects. Results are shown for system sizes $L=12$, $14$, and $16$.}
  \label{fig:OTOC_longtime}

\end{figure*}
  \begin{figure}[htbp]
  	\centering
  	\includegraphics[width=0.5\textwidth]{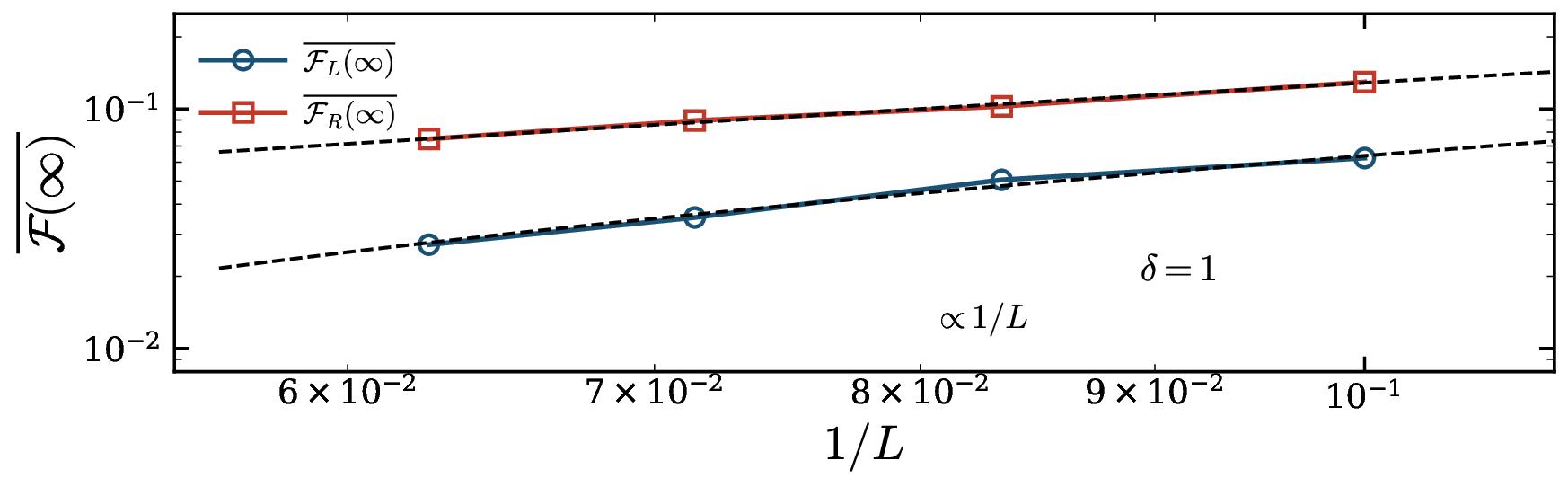}
  	\caption{(a) Finite-size scaling of the saturation value of the OTOC for the bond operator $S_i^zS_{i+1}^z$. The black dashed line shows the fit to the scaling relation $\mathcal{F}(\infty)\propto 1/L$. The results are shown for system sizes $L=10$, $12$, $14$, and $16$.}
  \label{fig:OTOC_longtime_scaling}
\end{figure}
The finite-size saturation value of OTOCs can be related to the overlap between the local operators and the Hamiltonian \cite{Balachandran_entropy,analytical2}. For two local unitary operators $W$ and $V$, the late-time saturation value is given by

\begin{equation}
	\begin{aligned}
		F_{L}^{WV}(\infty)
		&\approx
		\frac{1}{L}
		\frac{\langle WW^\dagger\rangle\,|\langle HV\rangle|^2 + \langle VV^\dagger\rangle\,|\langle HW\rangle|^2}{\langle HH_j\rangle }
		\\
		&\approx
		\frac{1}{L}\frac{|\langle HV\rangle|^2	+	|\langle HW\rangle|^2}{\langle HH_j\rangle},
	\end{aligned}
	\label{eq:PRL_result}
\end{equation}

    where $\langle\cdots\rangle=\mathrm{Tr}(\cdots)/ \mathcal D$ denotes the infinite-temperature     expectation value and $H_j$ is the local component of the Hamiltonian as $H = \sum_j H_j$, 
	$H_j =J_{xy}\left(S_j^xS_{j+1}^x+S_j^yS_{j+1}^y\right) +J_z(j;\delta)S_j^zS_{j+1}^z$. The locality of $H_i$ gives $\langle HH_j\rangle\simeq\langle H_j^2\rangle$, with terms decaying faster than $1/L$ neglected. For the analytical estimate, \(\langle H_j^2\rangle\) is averaged over all sites of the chain. For the bond operator $A_i=S_i^zS_{i+1}^z$, the overlap with the Hamiltonian is $\langle HA_i\rangle = J_z(i;\delta)$. For the left and right operators, the interaction parameters from Eq.~(\ref{eq:Jz_profile}) are $J_z(2;\delta)=1-\delta\frac{L-4}{L-2}$ and $J_z(L-2;\delta)=1+\delta\frac{L-4}{L-2}$, respectively, while for the center operator $J_z(L/2;\delta)=1$. Substituting these overlaps into Eq.~(\ref{eq:PRL_result}).


One can also understand the asymmetry in saturation of OTOCs by observing the distribution of diagonal elements of OTOC operators in the energy eigenbasis. The contribution from the product term in Eq.~(\ref{eq:FWV_inf}) is subleading for large system sizes~\cite{Balachandran2023}. Consequently, the infinite-time OTOC is well approximated by
\begin{equation}
	\mathcal{F}_{WV}(\infty)\approx
	\frac{1}{\mathcal D}
	\sum_{\alpha}
	\left(
	W_{\alpha\alpha}^{2}
	+
	V_{\alpha\alpha}^{2}
	\right).
	\label{eq:FWV_correction}
\end{equation}

 Equation~(\ref{eq:FWV_correction}) shows that the diagonal matrix elements of the local operators in the energy eigenbasis govern the long-time saturation value. Consequently, any spatial dependence of the diagonal matrix elements is directly reflected in the long-time OTOC.
 For the bond operator $S_i^zS_{i+1}^z,$ the slope of the diagonal matrix elements increases systematically from the weakly interacting side toward the strongly interacting side of the chain, as discussed in Appendix~\ref{app:diagonal_NN}. Consequently, the quantity $\sum_{\alpha}W_{\alpha\alpha}^2$ entering Eq.~(\ref{eq:FWV_correction}) is larger for the right probe than for the left probe, leading to the larger long-time saturation value observed in Fig.~\ref{fig:OTOC_longtime}. 

Figure~\ref{fig:OTOC_longtime} compares the numerical results with Eq.~(\ref{eq:PRL_result}) for $L=12$, $14$, and $16$. The approximate analytical expression captures the overall behavior observed in the numerical results. The deviations between the numerical data and the analytical prediction can be attributed to finite-size effects, and the agreement is expected to improve for larger system sizes.

Figure~\ref{fig:OTOC_longtime_scaling} shows the finite-size scaling of the long-time saturation value of the OTOC for the bond operator.

\section{Conclusion}
\label{sec:summary}

We have investigated the interplay between deterministic spatial inhomogeneity, eigenstate thermalization, and information scrambling in a disorder-free XXZ spin chain with spatially varying interactions. Spectral statistics show that increasing the interaction gradient drives the system from the integrable to a quantum-chaotic regime. In contrast, sufficiently large interaction gradients eventually drive the system toward a regime with suppressed level repulsion [see Figs.~\ref{fig:level_statistics}(a) and \ref{fig:level_statistics}(b)]. These results demonstrate that deterministic interaction gradients can drive the system from an integrable regime through a quantum-chaotic regime and, for sufficiently strong inhomogeneity, toward a regime characterized by suppressed level repulsion and inhibited thermalization. Although the conventional spectral diagnostics successfully identify the global crossover between these regimes, they are insensitive to the spatial structure of the interaction gradient. In contrast, OTOCs directly probe the spatial inhomogeneity by revealing a pronounced left--right asymmetry in information scrambling, which persists even within the quantum-chaotic regime.
To characterize information scrambling in the considered system (see Eq.\ref{eq:H_total}), we analyzed OTOCs and found that the interaction gradient induces a pronounced left--right asymmetry in the information scrambling. In particular, operators in the strongly interacting region exhibit suppressed scrambling, leading to larger finite-size saturation values.
To understand the relaxation dynamics of the OTOCs, we considered two classes of operators: onsite spin operators, which have vanishing overlap with the Hamiltonian, and bond operators, which possess a finite overlap. Bond operators exhibit slow algebraic relaxation, consistent with their finite operator--Hamiltonian overlap. In contrast, onsite operators display relatively rapid relaxation for weak interaction gradients. However, as the interaction gradient increases, they also develop slow relaxation despite having zero overlap with the Hamiltonian. These results demonstrate that spatial inhomogeneity significantly modifies the relaxation dynamics of information scrambling, suggesting that the operator--Hamiltonian overlap approach predicts only the fastest possible relaxation. In contrast, the actual relaxation can be substantially slower.
We further demonstrated that, in a stepwise interaction profile (see Fig. \ref{fig:stepwise_interaction}), OTOCs associated with probe regions with the same local interaction strength exhibit nearly identical dynamics, despite substantial inhomogeneity elsewhere in the chain. In contrast, a sufficiently large difference in the local interaction strengths produces pronounced asymmetric scrambling, with faster scrambling in the weakly interacting region and suppressed scrambling in the strongly interacting region .

We also investigated the finite-size long-time saturation values of the OTOCs using two approaches: (i) the first is based on the structure of the diagonal matrix elements of the OTOC observables in the energy eigenbasis, and (ii) the second employs a theoretical expression derived from the overlap between the Hamiltonian and the OTOC observables. The theoretical predictions are consistent with the numerical results (see Fig.~\ref{fig:OTOC_longtime}) and provide a microscopic understanding of the asymmetric finite-size saturation values.
Our results establish deterministic interaction gradients as an effective mechanism for controlling information scrambling in quantum spin chains. They further reveal how spatial inhomogeneity modifies both the relaxation dynamics and the finite-size long-time saturation of OTOCs.
\begin{acknowledgments}
	S.M. acknowledges financial support through a Senior Research Fellowship from Motilal Nehru National Institute of Technology (MNNIT) Prayagraj, India. S.M. also thanks Rohit Kumar Shukla for helpful conceptual discussions.
\end{acknowledgments}

\section*{Data Availability}
The data supporting the findings of this study are not publicly available but are available from the corresponding author upon reasonable request.

\appendix

\section{ OTOC Dynamics in an Inhomogeneous Floquet Chain}
\label{app:floquet_system}
\begin{figure*}[t]
	\centering
	\includegraphics[width=\textwidth]{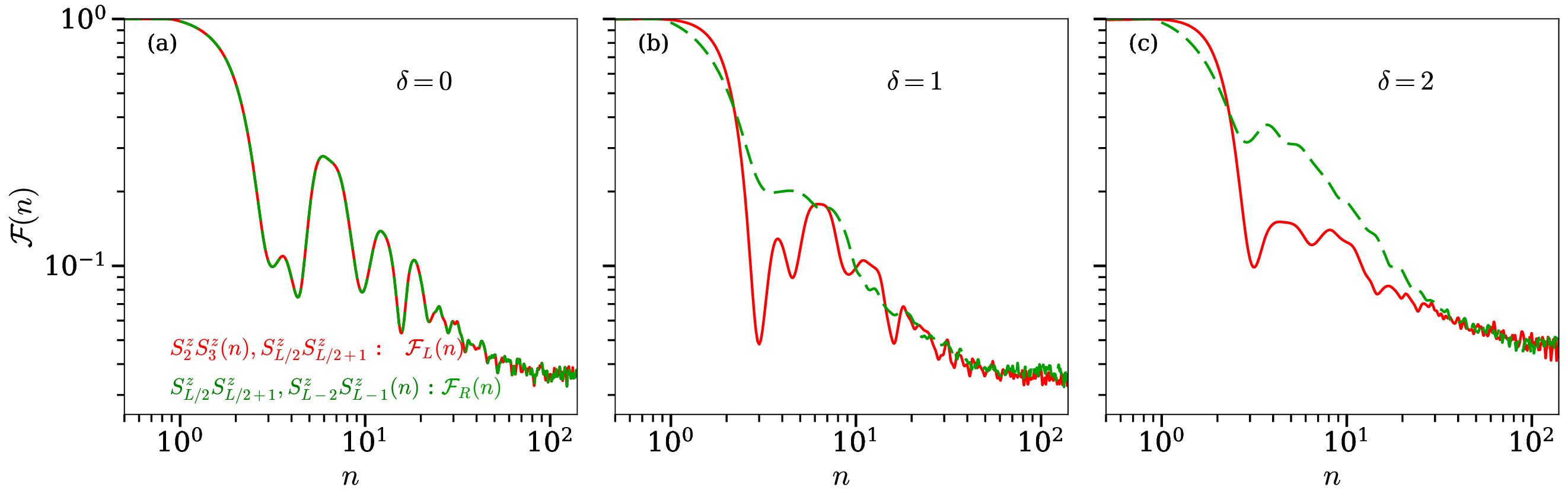}
	\caption{OTOC dynamics in the inhomogeneous Floquet XXZ chain for $\delta=0$, $1$, and $2$ in panels (a), (b), and (c), respectively. The solid red and dashed green curves represent the OTOCs $\mathcal{F}_{L}(n)$ and $\mathcal{F}_{R}(n)$ for probes located on the left and right sides of the chain, respectively. For the homogeneous case $\delta=0$, the OTOCs exhibit identical dynamics. As the interaction gradient increases, the left--right asymmetry in the OTOC dynamics becomes more pronounced, with the weakly interacting side exhibiting faster scrambling than the strongly interacting side. Results are shown for system size $L$=12.}
	\label{fig:floquet_fig}
\end{figure*}
To examine whether the asymmetric information scrambling observed in the time-independent inhomogeneous XXZ chain persists in a periodically driven system, we consider an inhomogeneous Floquet XXZ chain. The Floquet evolution over one driving period $T$ is defined as
\begin{equation}
	U_F(\delta)= e^{-iH_2(\delta)T/2}e^{-iH_1T/2},
	\label{eq:floquet_operator}
\end{equation}
where
\begin{align}
	H_1 &= J_{xy}\sum_{i=1}^{L-1}
	\left(S_i^xS_{i+1}^x+S_i^yS_{i+1}^y\right),\\
	H_2(\delta) &= \sum_{i=1}^{L-1}
	J_z(i;\delta)S_i^zS_{i+1}^z.
	\label{eq:floquet_H}
\end{align}
We characterize the scrambling dynamics through the infinite-temperature OTOC $\mathcal{F}_{WV}(n)=\frac{1}{D}\mathrm{Tr}W(n)VW(n)V$, the discrete-time Heisenberg evolution of an operator $W\equiv W(0)$ is given by $W(n)=U_F^{\dagger n}W(0)U_F^n$,
where $U_F$ is the Floquet evolution operator over one driving period. We set the driving period to $T=1$. In the homogeneous limit ($\delta=0$), the left and right OTOCs coincide. For finite $\delta$, the interaction strength increases from left to right, resulting in inequivalent local scrambling dynamics. In particular, the OTOC associated with the strongly interacting region exhibits slow scrambling compared with that on the weakly interacting side. Thus, the Floquet dynamics provide an independent demonstration that spatially varying interactions can generate asymmetric information scrambling in periodically driven quantum many-body systems (see Fig. \ref{fig:floquet_fig}). The diagonal matrix elements of local operators in the eigenbasis of the Floquet system are independent of quasienergy and largely structureless (see, e.g., Ref.~\cite{floquet}), in contrast to those of the static XXZ Hamiltonian. Consequently, the interaction-gradient-induced asymmetry in the long-time saturation values is not observed in the Floquet system. 

in contrast to the even-size case, where the diagonal matrix elements of $S_i^z$ vanish in the zero-magnetization sector as a consequence of the spin-flip symmetry. 


\section{OTOCs in the Inhomogeneous Transverse-Field Ising Model (ITFIM)}
\label{app:itfim_otoc}

To demonstrate that the asymmetric information scrambling discussed in the main text is not unique to the inhomogeneous XXZ chain, we consider a transverse-field Ising model (TFIM) \cite{tfim1,tfim2} with a spatially varying transverse field, referred to here as the inhomogeneous transverse-field Ising model (ITFIM), described by
\begin{equation}
	H=
	h_z\sum_{i=1}^{L}S_i^z
	+h_x\sum_{i=1}^{L}S_i^x
	+\sum_{i=1}^{L-1}
	J_z(i;\delta)\,
	S_i^zS_{i+1}^z,
	\label{eq:ITFIM}
\end{equation}
where the nearest-neighbor Ising interaction varies spatially according to $J_z(i;\delta)=J_0+\delta\frac{2i-L}{L-2}$,	with $\delta$ controlling the strength of the interaction gradient.	We set Hamiltonian parameters as $h_x=0.9$, $h_z=0.809$, and $J_0=1.0$.

\begin{figure}[t]
	\centering
	\includegraphics[width=0.48\textwidth]{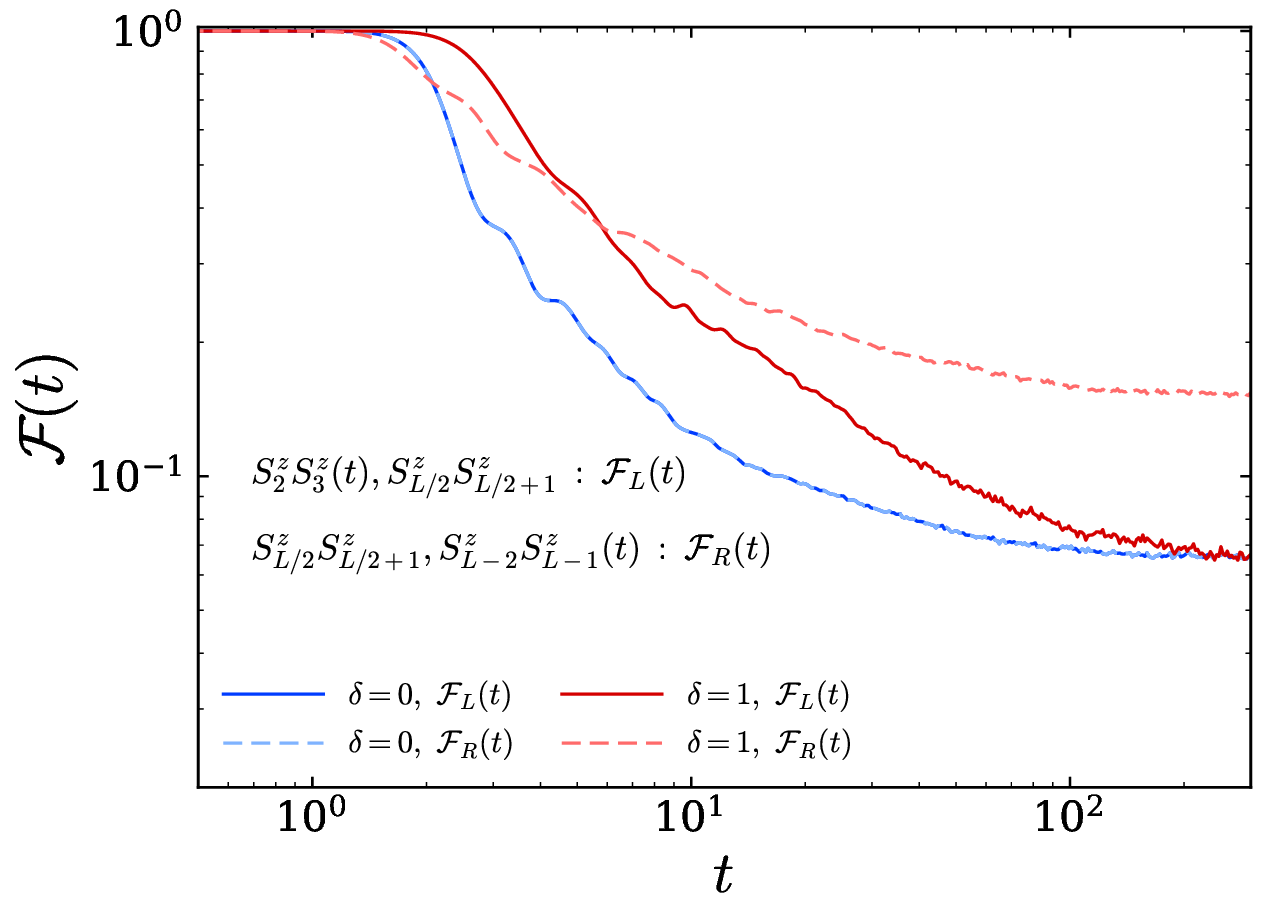}
	\caption{Dynamics of OTOCs for the ITFIM with interaction-gradient strengths $\delta=0$ and $1$ for system size $L = 12$. The left probe, $\mathcal{F}_{L}(t)$, is constructed from bond operators near the weakly interacting edge, $S_{2}^{z}S_{3}^{z}$ and $S_{L/2-1}^{z}S_{L/2}^{z}$, while the right probe, $\mathcal{F}_{R}(t)$, uses the corresponding bond operators near the strongly interacting edge, $S_{L/2}^{z}S_{L/2+1}^{z}$ and $S_{L-2}^{z}S_{L-1}^{z}$. In the homogeneous limit ($\delta=0$), the two OTOCs coincide. Increasing the interaction gradient produces a clear spatial asymmetry: the strongly interacting side exhibits suppressed scrambling, characterized by slower OTOC decay than on the weakly interacting side.}
	\label{fig:itfim_otoc}
\end{figure}

Figure~\ref{fig:itfim_otoc} presents the resulting OTOCs for several values of the interaction-gradient strength $\delta$. Similar to the inhomogeneous XXZ chain, the ITFIM exhibits a pronounced left--right asymmetry in the scrambling information. Increasing the interaction gradient suppresses operator spreading in the strongly interacting region, leading to suppression of scrambling than observed on the weakly interacting side. These results demonstrate that asymmetric information scrambling is a generic consequence of deterministic spatial inhomogeneity rather than a feature specific to the XXZ model.

\subsection{Diagonal Matrix Elements}
\label{app:diagonal_NN}
\begin{figure}[t]
	\centering
	\includegraphics[width=0.48\textwidth]{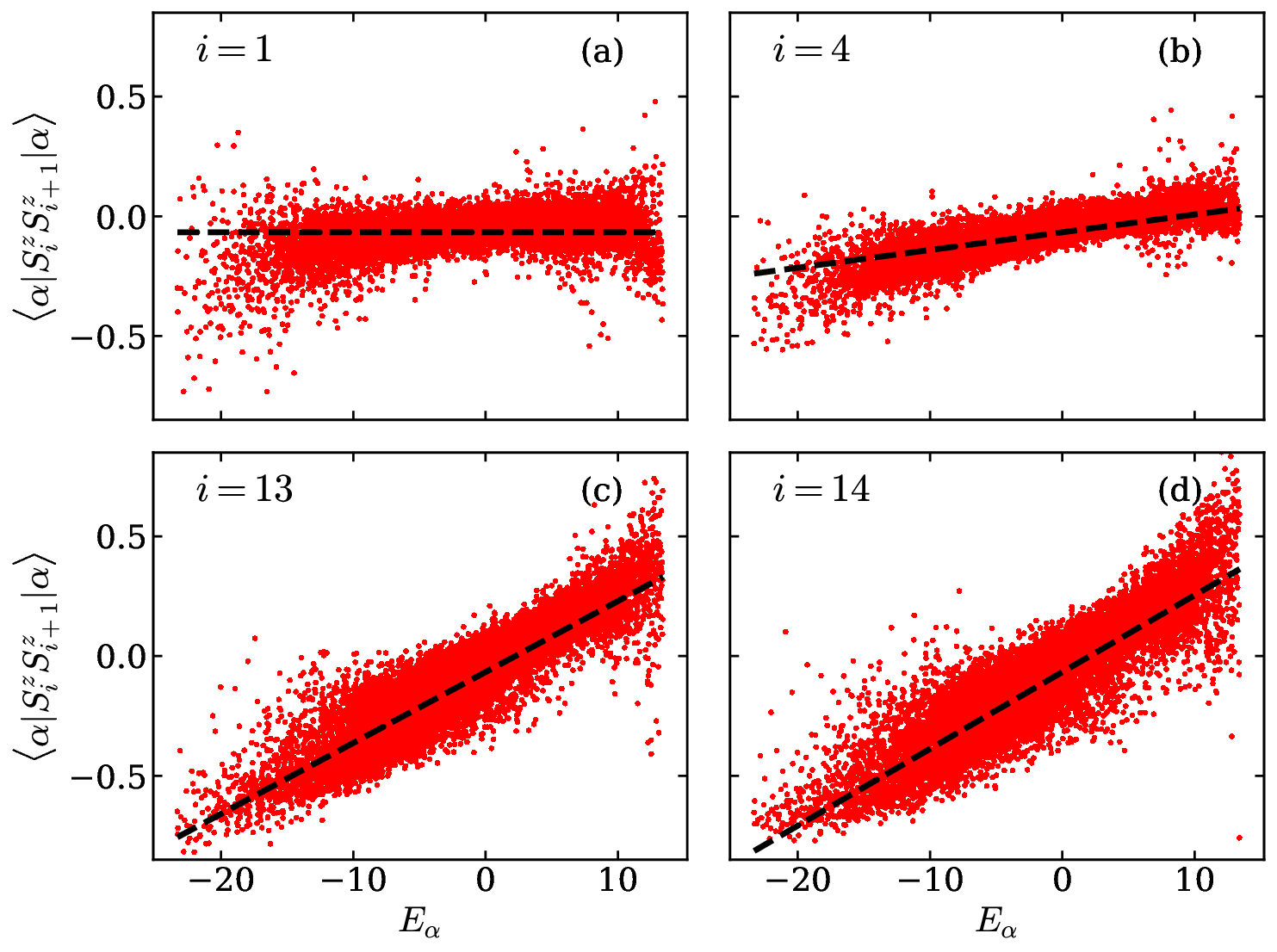}
	\caption{(a)--(d) Diagonal ETH matrix elements $\langle\alpha|S_i^zS_{i+1}^z|\alpha\rangle$ for the bond operator at $i=1$, $4$, $13$, and $14$, respectively, plotted as functions of the eigenenergy $E_\alpha$. Results are shown for the inhomogeneous XXZ chain with $L=16$ and $\delta=1$. The black dashed lines represent linear fits to the analytical prediction of Eq.~(\ref{eq:analytical_bond}), demonstrating that the diagonal matrix elements vary linearly with the eigenenergy.}
	\label{fig:analytical_eth}
\end{figure}

Within the framework of the ETH, the matrix elements of a local observable $O$ in the eigenbasis of the Hamiltonian $H$ are expressed as \cite{5,16,17,18,20}
\begin{align}
	\langle \alpha | O | \beta \rangle	=\bar{O}(\bar{E})\delta_{\alpha\beta} +	e^{-S(\bar{E})/2}
	f(\bar{E},\omega) R_{\alpha\beta},
	\label{Eq:ETH}
\end{align}
where $\bar{E}=(E_\alpha+E_\beta)/2$, $\omega=E_\alpha-E_\beta$, $S(\bar{E})$ is the thermodynamic entropy, $\bar{O}$ and $f$ are smooth functions, and $R_{\alpha\beta}$ is a random variable with zero mean and unit variance. In chaotic systems, the diagonal matrix elements vary smoothly with energy and the off-diagonal elements follow Gaussian statistics. By contrast, integrable systems exhibit strong eigenstate-to-eigenstate fluctuations and non-Gaussian off-diagonal matrix elements, thereby violating ETH \cite{21,22,23,25,mishra2026,27,28}.
The diagonal ansatz of ETH is given as
\begin{equation}
	O_{\alpha \alpha} = O(E_{\alpha}) + e^{-S(E_{\alpha}) / 2} f_O(E_{\alpha}, 0) R_{\alpha \alpha},
\end{equation}
Since the entropy is an extensive quantity, the second term decreases exponentially with the system size. 
We can also write
\begin{equation}
	\left| \langle \alpha | O | \alpha \rangle - f_O(E_\alpha/L) \right| \le e^{-\Omega(L)},
	\label{eq:eth_ansatz}
\end{equation}
The function $f_O(E_\alpha/L)$ can be expanded as
\begin{equation}
	f_O(E_\alpha/L) = \sum_{q=0}^{\infty} \frac{f_O^{(q)}(0)}{q!} \left( \frac{E_\alpha}{L} \right)^q,
\end{equation}
with $f_O^{(q)}(0)$ being the $q$-th derivative. The moments of the Hamiltonian satisfy \cite{analytical2}
\begin{equation}
	\frac{1}{\mathcal{D}} \sum_\alpha E_\alpha^q = \langle H^q \rangle = \mathcal O(L^{q/2}),
	\label{eq:H_square}
\end{equation}
where $\mathcal{D}$ is the Hilbert space dimension of the $S^z_{\text{tot}}=0$ sector.

For traceless local observables,  $f_O(0)=\frac{\operatorname{Tr}(O)}{\mathcal{D}}$ is zero. Furthermore, if the first derivative satisfies
$f_O^{(1)}(0)\neq0$, then one obtains \cite{analytical1, Balachandran_entropy}
\begin{align}
	\operatorname{Tr}(OH)
	&=
	\frac{1}{\mathcal{D}}
	\sum_{\alpha}
	O_{\alpha\alpha}E_{\alpha} \simeq \frac{1}{\mathcal{D}L} \sum_{\alpha}	E_{\alpha}^{2}\,
	f_O^{(1)}(0)
	\nonumber\\
	&\approx
	\frac{\langle H^{2}\rangle}{L}
	\,f_O^{(1)}(0).
	\label{eq:traceOH}
\end{align}

which immediately gives
\begin{equation}
	f_O^{(1)}(0)
	\simeq
	\frac{\operatorname{Tr}(OH)\,L}
	{\mathcal D\,\langle H^2\rangle},
	\label{eq:firstderivative} 	
\end{equation}
where $\langle H^2\rangle \propto L$ from Eq. \ref{eq:H_square}.
In the $S_{\mathrm{tot}}^z=0$ sector, the bond operator $A_i=S_i^zS_{i+1}^z$ satisfies $f_{A_i}(0)\neq0$. Since $f_{A_i}(0)=-\dfrac{1}{(L-1)}$, the leading energy-dependent contribution to the smooth ETH function is determined by the first derivative $f_{A_i}^{(1)}(0)$.  Using the relation $f_{A_i}^{(1)}(0)\propto \operatorname{Tr}(A_iH)L/(\mathcal D\,\langle H^2\rangle)$, we obtain $f_{A_i}^{(1)}(0) \propto J_z(i;\delta)$. Retaining the leading term in the Taylor expansion gives
\begin{equation}
	\langle \alpha | S_i^z S_{i+1}^z | \alpha \rangle \propto J_z(i;\delta) \, \left( \frac{E_\alpha}{L} \right) -\frac{1}{(L-1)}.
	\label{eq:analytical_bond}
\end{equation}

The diagonal matrix elements for the bond operator $S_i^zS_{i+1}^z$, together with the corresponding microcanonical line, are shown in Fig.~\ref{fig:analytical_eth}. The slope of the microcanonical line is proportional to the local interaction strength $J_z(i;\delta)$. Consequently, the distribution of the diagonal matrix elements rotates systematically as the interaction gradient increases. These results demonstrate how spatial inhomogeneity influences the structure of the diagonal matrix elements.

\newpage

\bibliography{ref}

\end{document}